\documentclass[12pt]{article}
\pdfoutput=1
\usepackage{amsmath,amssymb,graphicx}
\usepackage{hyperref}
\usepackage[T1]{fontenc}
\usepackage[charter]{mathdesign}
\usepackage[font=small,labelfont=bf]{caption}
\usepackage{cancel}
\usepackage{cite}
\usepackage{booktabs}

\newcommand{\beq}{\begin{equation}}
\newcommand{\eeq}{\end{equation}}
\newcommand{\beqs}{\begin{eqnarray}}
\newcommand{\eeqs}{\end{eqnarray}}
\def\simlt{\stackrel{<}{{}_\sim}}
\def\simgt{\stackrel{>}{{}_\sim}}

\usepackage[usenames,dvipsnames]{xcolor}
\usepackage{framed}

\title{\bf \color{blue!20!gray} Single-Scale Natural SUSY}
\author{Lisa Randall and Matthew Reece \\
{\small \texttt{randall, mreece@physics.harvard.edu}}\\
\em{Department of Physics, Harvard University, Cambridge, MA 02138}}

\begin{document}
\maketitle

\begin{abstract}
We consider the prospects for natural SUSY models consistent with current data. Recent constraints make the standard paradigm unnatural so we consider what could be a minimal extension consistent with what we now know. The most promising such scenarios  extend the MSSM with new tree-level Higgs interactions that can lift its mass to at least 125 GeV and also allow for flavor-dependent soft terms so that the third generation squarks are lighter than current bounds on the first and second generation squarks. We argue that a common feature of almost all such models is the need for a new scale near 10 TeV, such as a scale of Higgsing or confinement of a new gauge group. We consider the question whether such a model can naturally derive from a single mass scale associated with supersymmetry breaking.  Most such models simply postulate new scales, leaving their proximity to the scale of MSSM soft terms a mystery. This coincidence problem may be thought of as a mild tuning, analogous to the usual $\mu$ problem. We find that a single mass scale origin is challenging, but suggest that a more natural origin for such a new dynamical scale is the gravitino mass, $m_{3/2}$, in theories where the MSSM soft terms are a loop factor below $m_{3/2}$. As an example, we build a variant of the NMSSM where the singlet $S$ is composite, and the strong dynamics leading to compositeness is triggered by masses of order $m_{3/2}$ for some fields. Our focus is the Higgs sector, but our model is compatible with a light stop (either with the first and second generation squarks heavy, or with $R$-parity violation or another mechanism to hide them from current searches). All the interesting low-energy mass scales, including linear terms for $S$ playing a key role in EWSB, arise dynamically from the single scale $m_{3/2}$. However, numerical coefficients from RG effects and wavefunction factors in an extra dimension complicate the otherwise simple story.
\end{abstract}

\section{The State of SUSY: Introduction}
\label{sec:intro}

The Large Hadron Collider (LHC) has recently made significant progress toward one of its central physics goals: understanding the nature of electroweak symmetry breaking. In nearly 5 fb$^{-1}$ of data collected in 2011, both ATLAS and CMS have observed hints of an approximately Standard Model-like Higgs boson with a mass near 125 GeV~\cite{ATLAS:2012ae,Chatrchyan:2012tx,Pieri:2012sb,Bernhard:2012be}. Since the initial preprint of this paper, data taken at 8 TeV have confirmed the discovery of a new boson~\cite{Aad:2012tfa,Chatrchyan:2012ufa}. Future measurements will continue to probe the couplings of this particle. Given that the complete absence of signals of the higher-dimension operators that could portend a strongly-coupled explanation of EWSB had already made our prior expectation for a Higgs-like explanation very high, and given current data, it appears that the discovered particle couples in a way very similar to Standard Model predictions for the Higgs boson. The measured Higgs mass is near the high end of the mass range that would be expected for minimal supersymmetry.

A 125 GeV Higgs boson with approximately SM-like couplings reinforces the weak hierarchy problem: how does the large ratio of the Planck scale to the weak scale persist in light of quantum mechanical effects? The natural options essentially fall into two categories. The first is that the Higgs is a composite state, so that the cutoff scale is nearby. If the Higgs were a typical composite state, we would expect other states to have been observed as well, at least indirectly. It may be an accidentally light state, as in Randall-Sundrum theories~\cite{Randall:1999ee}. Or, it could be a pseudo-Nambu-Goldstone boson, with a radiatively generated potential~\cite{Kaplan:1983fs,Kaplan:1983sm,Georgi:1984ef,Contino:2003ve,Agashe:2004rs}. In this case, tuning is required to achieve a Higgs VEV much less than the pion decay constant in the strongly interacting sector; after such a tuning, Higgs masses of 125 GeV, or significantly heavier, can be accommodated. The second option is supersymmetry, the only known mechanism allowing truly elementary scalars to be natural. This will be our focus in this paper.

Indirect measurements showing the absence of flavor-changing neutral currents~\cite{Isidori:2010kg} and electric dipole moments~\cite{Regan:2002ta,Griffith:2009zz} put important constraints on supersymmetry, but could be avoided by sufficiently symmetric models of supersymmetry breaking. Direct searches at the LHC have now highly constrained even that option, putting supersymmetry in an awkward position. The jets plus missing energy signatures that are generally considered its hallmark have not been found in 7 TeV data, putting bounds of above 1 TeV on squarks and gluinos decaying through light-flavor jets~\cite{Aad:2011ib,CMS-PAS-SUS-11-004,ATLAS-CONF-2012-033,CMS-PAS-SUS-12-011} and (adding leptons or $b$-jets to the search) a slightly weaker bound on a gluino decaying through the third generation~\cite{ATLAS-CONF-2012-004,Collaboration:2012sa,CMS-PAS-SUS-11-027,ATLAS-CONF-2012-037,ATLAS-CONF-2012-058}. More recently, analyses of 8 TeV data have increased the bound on gluinos decaying through third generation quarks to around 1.3 TeV~\cite{ATLAS-CONF-2012-145,Chatrchyan:2013lya}. Furthermore, direct searches for sbottom and stop production have been published, ruling out (for example) sbottom pair production with ${\tilde b}_1 \to b{\tilde \chi}^0_1$ for sbottom masses below about 650 GeV and very light neutralinos, and stop pair production with ${\tilde t}_1 \to t{\tilde \chi}^0_1$ for stop masses between about 250 and 650 GeV and very light neutralinos~\cite{Chatrchyan:2013lya,Chatrchyan:2012wa,CMS-PAS-SUS-12-023,ATLAS-CONF-2012-166,ATLAS-CONF-2013-001,ATLAS-CONF-2013-007,ATLAS-CONF-2013-024}. In every case, the limits degrade when the LSP is made heavier, often disappearing completely for LSP masses above 200 or 300 GeV.

These direct bounds on superpartners begin to threaten supersymmetric naturalness, and should be confronted with the expected ``naturalness bounds'' on masses assuming low fine-tuning, as discussed recently in Refs.~\cite{Papucci:2011wy,Brust:2011tb}. Allowing a 10\% tuning, one finds roughly that the gluino mass should be below about 1.3 TeV and the root-mean-square stop mass below about 800 GeV. While a large fraction of this parameter space has been ruled out, some of it still remains. For instance, stops near the top mass are still allowed by data, and bounds on all squarks and gluinos become weaker as the lightest neutralino mass is raised. Direct searches can also be evaded with models that modify decay chains. Hence, while the direct searches offer no reassuring indications of natural physics, they have not yet ruled it out. The measured mass of the Higgs boson is more troubling in the MSSM than the direct bounds on SUSY. A Higgs mass of 125 GeV, because the tree-level Higgs quartic is related to the electroweak gauge couplings, requires large loop corrections from heavy stops or large $A$-terms. This necessitates a high degree of fine-tuning in the MSSM, of order a part in a few hundred to a part in a thousand or more, and all but excludes large classes of models that were previously plausible~\cite{Hall:2011aa,Baer:2011ab,Heinemeyer:2011aa,Arbey:2011ab,Draper:2011aa,Carena:2011aa}.

If supersymmetry is to play a role in stabilizing the hierarchy, we are left with a dilemma. On the one hand, we can continue to study the MSSM as a possible answer, weakening our requirement of naturalness to accommodate some amount of fine-tuning. For instance, the stops could be at 10 TeV, and supersymmetry could explain the large hierarchy between this scale and the Planck scale, leaving the little hierarchy between the weak scale and 10 TeV unexplained.  On the other hand, we can insist that naturalness remain a strong guiding principle, in which case the stop squarks must be light to cancel large divergent contributions to $m_{H_u}^2$. In this case, the Higgs mass becomes the difficulty, and we must look for physics beyond the MSSM to explain how it came to be at 125 GeV rather than near 90 GeV. We would also like such a theory to predict, or at least accommodate, a flavored superpartner spectrum, so that the stops and sbottoms can remain relatively light while the first and second generation squarks can be safely heavy enough to avoid constraints~\cite{Dimopoulos:1995mi,Cohen:1996vb}. Because the tree-level contributions to the Higgs potential can be significantly larger in a theory beyond the MSSM, the stops can be heavier at fixed tuning measure~\cite{Barbieri:1987fn}, even reaching 1.4 TeV with only 10\% tuning in some scenarios~\cite{Hall:2011aa}. However, bounds on first- and second-generation squarks already exclude such masses (at least for typical $R$-parity conserving decays)~\cite{ATLAS-CONF-2012-033}, so the data suggest that a natural SUSY model should have generation-dependent soft terms. Alternatively, the squarks could have degenerate soft masses, evading the SUSY flavor problem, but could have evaded detection so far by decaying, as in $R$-parity violating models~\cite{Barbier:2004ez,Csaki:2011ge,Graham:2012th}.

In this paper we highlight a common thorny model-building issue in theories that extend the MSSM to produce a 125 GeV Higgs: they typically require a scale near the TeV scale (often at about 10 TeV) that is a priori unrelated to the scale of SUSY-breaking soft masses. This may be thought of as an additional (often logarithmic) tuning that such theories require, which weakens their appeal over the finely tuned MSSM. We are thus motivated to construct ``single-scale'' natural SUSY models, in which no accidental coincidence of scales is required. The essential idea is that two scales, $m_{3/2}$ and $\frac{g^2}{16\pi^2} m_{3/2}$, can arise from one SUSY-breaking parameter, so that single-scale natural SUSY works very well with scenarios with $m_{3/2} \sim 100$ TeV.

In the next section, we review the basic approaches to raising the Higgs mass through nondecoupling $D$- or $F$-terms, and explain why they typically require a new mass scale below 10 TeV. We also briefly review how natural SUSY models can separate the first and second generation soft masses from the third. In Section~\ref{sec:singlescale}, we construct a more elaborate example of an NMSSM-like theory with a composite singlet $S$, similar to that of Ref.~\cite{Chang:2004db}, but with the compositeness scale and other scales in the superpotential determined by $m_{3/2}$. In particular, loops in this model generate an effective $fS$ superpotential term as well a SUSY-breaking $S$ tadpole, making it much easier to achieve electroweak symmetry breaking than in more traditional NMSSM-like theories where a large negative mass squared for $S$ is required. This suggests that a single-scale natural SUSY Higgs sector can be achieved. We offer brief remarks on the flavor sector of such a theory (which should have either split generations or a mechanism like $R$-parity violation to hide squarks), but defer a detailed analysis for future work, because it is orthogonal to the Higgs sector modeling that is our main focus. One less attractive feature of the details of the specific scenario we propose for the Higgs sector is that it necessarily contains large and small numerical factors from renormalization-group effects of strong dynamics and wavefunction overlaps in an extra dimension, which complicate the parametric simplicity of relying on one scale. We offer some concluding remarks in Section~\ref{sec:discuss}.

\section{The trouble with models}
\label{sec:models}

\subsection{New quartics generically demand a new scale below about 10 TeV}

Our goal in this section is to briefly review mechanisms for explaining a Higgs mass of 125 GeV in natural SUSY models (see also~\cite{Lodone:2012kp}), and show that they usually require a new mass scale near 10 TeV.

The common feature of models of natural SUSY compatible with experimental constraints is that they provide new contributions to the Higgs quartic. The difficulty is that in the MSSM, corrections to the Higgs/$Z$ mass relationship~\cite{Haber:1990aw,Barbieri:1990ja,Carena:2002es} are only logarithmic,
\beq
m_h^2 \approx m_Z^2 \cos^2 2\beta + \frac{3 m_t^4}{4\pi^2 v^2} \left(\log \frac{m_{\tilde t}^2}{m_t^2} + \frac{X_t^2}{m_{\tilde t}^2} \left(1 - \frac{X_t^2}{12 m_{\tilde t}^2}\right)\right),
\eeq
with $X_t$ the left/right stop mixing parameter $A_t - \mu \cot\beta$, whereas the RGE
\beq
\frac{d}{dt} m_{H_u}^2 = \frac{3 y_t^2}{8\pi^2} \left(m_{Q_3}^2 + m_{u^c_3}^2 + m_{H_u}^2 + A_t^2\right)
\eeq
shows that the corrections to the soft mass of the $H_u$ multiplet are quadratic in superpartner masses and the $A$-term. This implies that in the MSSM, a Higgs significantly heavier than the $Z$ will imply sufficiently large quadratic corrections to require fine tuning for electroweak symmetry breaking. Furthermore, any additional matter introduced to raise the Higgs mass through loops~\cite{Babu:2008ge,Martin:2009bg,Graham:2009gy} will incur a similar (but perhaps numerically smaller) tuning cost.

In other words, natural SUSY demands new {\em tree-level} quartic couplings for the Higgs~\cite{Casas:2003jx}. Furthermore, we argue that these quartics necessarily involve new physics with a mass scale not far above the TeV scale. As a first class of examples, let us consider new quartics that arise from $D$-term potentials associated with gauge symmetries~\cite{Haber:1986gz,Randall:2002talk,Batra:2003nj,Maloney:2004rc,Craig:2011yk}. Any new symmetry under which the Higgs is charged will give a new quartic, but if we can integrate out the gauge boson supersymmetrically, the quartic $D$-term interaction will be canceled by the exchange of the heavy modes. As a result, the physical effect is proportional to soft masses of the scalars $\phi, {\bar \phi}$ that Higgs the heavy gauge boson; e.g., for a new U(1)$_x$ symmetry under which $H_u$ and $H_d$ have opposite charge $\pm 1$, one obtains a term~\cite{Batra:2003nj}
\beq
\delta V_{D-{\rm term}} = \frac{m_\phi^2}{M_{Z_x}^2 + 2 m_\phi^2} g_x^2 \left(\left|H_u\right|^2 - \left|H_d\right|^2\right)^2.
\eeq
This is an effective hard SUSY-breaking term. If the soft mass $m_\phi^2$ for the Higgsing of the new symmetry is of the same order as MSSM soft masses, which we take to be at or below 1 TeV, it is clear that obtaining a large effect from this term demands $M_{Z_x} \simlt 10$ TeV. A large gauge coupling $g_x$ doesn't help in that it raises the gauge boson mass as well. One can consider larger SUSY breaking in this U(1) sector, with $m_\phi^2 \simgt M_{Z_x}^2$, to approach a truly non-decoupling limit. However, the scale cannot be far above the TeV scale: the effective theory has hard SUSY breaking, so a quadratically divergent Higgs mass proportional to the new quartic, cut off at the scale $M_{Z_x}$, is generated. Naturalness then demands $M_{Z_x} \simlt 10$ TeV.

The other general category of models involves new $F$-term quartics~\cite{Drees:1988fc,Espinosa:1991gr}. The chief example is the NMSSM, which broadly construed encompasses theories that have an effective low-energy superpotential
\beq
W = \lambda S H_u H_d + f(S) + W_{MSSM}.
\eeq
In the most general case  all possible functions $f(S)$ are included as well as  a possible $\mu$-term in $W_{MSSM}$ not arising from $S$ (see the extensive review~\cite{Ellwanger:2009dp}). The tree-level Higgs quartic potential has a new term, $\left|F_S\right|^2 \supset \left|\lambda H_u H_d\right|^2$, which (involving both $H_u$ and $H_d$) becomes largest at {\em small} tan beta. This somewhat limits its efficacy, but it can improve the naturalness of a 125 GeV Higgs mass beyond that of the MSSM, especially when $\lambda$ is large and $\tan\beta \approx 2$~\cite{Hall:2011aa}.

This is fine as an effective theory. For $\lambda \simgt 0.7$, as is well-known, this theory does not remain perturbative up to the GUT scale. Even if we consider $\lambda$ small enough that the theory is valid well above the TeV scale, we encounter another obstacle. If the field $S$ is truly a singlet, it is allowed to have a tadpole. Such tadpoles are disastrous, completely destabilizing the hierarchy~\cite{Bagger:1993ji,Bagger:1995ay}. Planck-suppressed K\"ahler potential operators give rise, in supergravity, to {\em hard} SUSY-breaking terms in the Lagrangian like
\beq
\frac{c}{M_P} m_{3/2}^2 \left(S + S^\dagger\right) \left| H_u \right|^2 \Rightarrow~{\rm Tadpole:} \sim \frac{c}{16\pi^2 M_P} m_{3/2}^2 \Lambda^2 \left(S + S^\dagger\right).
\eeq
The quadratic divergence here makes this term dangerous; $S$ can get a VEV so large that it lifts the Higgs mass well above the TeV scale. One requires $m_{3/2} < 1$ keV to avoid this problem. But even in a theory of low-scale SUSY breaking, one can have other sources of tadpoles for $S$ that can be problematic. The only safe way to avoid destabilizing divergences is to charge $S$ under some symmetry. The traditional choice is a discrete ${\mathbb Z}/3$ symmetry under which $S$, $H_u$, and $H_d$ all have charge 1. However, this discrete symmetry leads to a cosmological domain wall problem; breaking the symmetry enough to have a safe cosmology reintroduces the tadpole problem~\cite{Abel:1995wk}. Sufficiently complicated discrete $R$-symmetry choices may alleviate this problem by forbidding $S$, $S^2$, {\em and} $S^3$ superpotential terms and attempting to generate an $S$ tadpole of the right size from a high-loop diagram~\cite{Panagiotakopoulos:1999ah}. Because the $R$-symmetry is broken at a scale $\sqrt{F} \gg~{\rm TeV}$, the domain wall problem may be avoided by a sufficiently low inflation scale. Such a model may constitute a loophole of our claim of generic new physics at 10 TeV, but it is a complicated one (see also the more recent work~\cite{Lee:2011dya,Ross:2011xv,Ross:2012nr}). Another way to avoid both the tadpole and the domain wall problems is to charge $S$ and the Higgs fields under a new U(1) symmetry broken near the TeV scale, combining aspects of the $D$- and $F$-term models~\cite{Cvetic:1997ky,Langacker:1999hs}.

Recently several groups have embraced the need for a low cutoff to remove destabilizing divergences, allowing consideration of large $\lambda \approx 2$ so that the effective theory hits a Landau pole at about 10 TeV. Such ``$\lambda$SUSY'' models~\cite{Barbieri:2006bg,Barbieri:2010pd} may be UV completed into a theory where one or more particles, including $S$, are composite~\cite{Harnik:2003rs,Chang:2004db,Delgado:2005fq,Craig:2011ev,Csaki:2011xn,Csaki:2012fh}. Such theories can be very natural from the standpoint of electroweak symmetry breaking. The model \cite{Csaki:2012fh} provides a natural framework for composites, a natural Higgs potential,  and a split spectrum. However one does need to assume mass scales are all roughly of the same order. The class of models of $\lambda$SUSY type provides a framework for studying the Higgs sector while remaining agnostic about the UV completion. (See also Refs.~\cite{Delgado: 2010uj,Delgado:2012yd} in a similar spirit.) With even less commitment to a UV completion, the lifting of the Higgs mass using higher dimension operators suppressed by a scale of several TeV has been studied~\cite{Dine:2007xi}. Such higher dimension operators can also arise when a natural SUSY effective theory emerges in the infrared of a strongly interacting theory~\cite{Gherghetta:2003he,Sundrum:2009gv}.

Models where the new $F$-terms arise from triplets often involve a singlet as well~\cite{Espinosa:1991gr,Agashe:2011ia}, in which case a new scale near 10 TeV is expected for the reasons already explained. Others require two opposite-hypercharge triplets with a mass term $M_T T \overline{T}$, in which case the mass scale $M_T$ plays a similar role. Another interesting approach involves adding superpotential operators coupling the Higgs fields to a new sector, $H_u {\cal O}_d + H_d {\cal O}_u$~\cite{Evans:2010ed,Azatov:2011ht,Azatov:2011ps,Gherghetta:2011na,Heckman:2011bb,Kitano:new}, where the strong dynamics associated with the ${\cal O}$ fields may play a role in EWSB.

\subsection{The tuning cost of coincident scales}
\label{sec:tuningscales}

We have seen that extensions of the MSSM that allow for $m_h \approx 125$ GeV typically have a new mass scale around 10 TeV or below. Broadly speaking, this is the scale of Higgsing a new gauge group in models with $D$-term quartics, and of compositeness in models with $F$-term quartics. In a supersymmetric theory, such a scale can always be technically natural. A superpotential interaction such as $X(\Phi_+ \Phi_- - f^2)$ given the nonrenormalization of the superpotential can generate the scale $f$ of Higgsing. In the case of compositeness, the 10 TeV scale can be just as natural as $\Lambda_{\rm QCD}$, arising from dimensional transmutation.  The problem in both cases is that the scales are unsatisfying,  since we need to assume a near coincidence of a new scale with the scale of supersymmetry breaking, which is in principle completely independent.

 Without further dynamical connection, there is therefore a  mild coincidence problem. The theory would clearly be more satisfying if the scales were tied together in a natural way. If enough scales were required to be coincident we might prefer the ordinary MSSM with heavy scalars, despite its fine-tuning.  Even for a  superpotential mass that isn't renormalized, despite technical naturalness, it seems very unlikely that UV physics would have the scale come out right. If the scale arises by dimensional transmutation, it is more appealing, since exponentially small numbers naturally arise and we are effectively adjusting the log of the scale instead of the scale itself. Even so, there is a coincidence we would prefer to explain as a consequence of dynamics.

A few examples already successfully relate the various new scales. A class of NMSSM-like models with a U(1) gauge symmetry under which $S$ is charged achieve radiative breaking of the U(1), relating its mass to the SUSY-breaking scale~\cite{Cvetic:1997ky,Langacker:1999hs}. For very low-scale SUSY breaking, operators suppressed by the messenger scale can lift the Higgs mass~\cite{Casas:2003jx}. Single-sector models assume that strong dynamics breaks supersymmetry at around 100 TeV, also producing composite first- and second-generation superparticles~\cite{ArkaniHamed:1997fq,Luty:1998vr,Franco:2009wf,Craig:2009hf,SchaferNameki:2010iz}. Another recent approach attempts to have the NMSSM on an IR brane in warped space, with the IR brane scale large and unrelated to supersymmetry breaking, and to have EWSB happen radiatively~\cite{Larsen:2012rq}. In such radiative NMSSM models, it is difficult to generate $m_s^2$ tachyonic enough for reasonable EWSB, given that $S$ is a singlet so interactions that can push it negative are typically weak~\cite{Agashe:1997kn, deGouvea:1997cx, Morrissey:2008gm}.

In this paper we ask if we can do better and how far we can go in the direction of a supersymmetric model consistent with all existing constraints and with naturalness. With this goal in mind---a more natural solution to the problem of coincidence of scales---as we will explore in the context of an example in Section~\ref{sec:singlescale}, we consider the possibility that all the scales in the problem arise from the supersymmetry breaking scale $m_{3/2}$. This works best in scenarios, like anomaly mediation~\cite{Randall:1998uk,Giudice:1998xp}, in which MSSM soft-breaking terms are a loop factor below $m_{3/2}$, which is then near 30 TeV. Although not our primary motivation, an independent reason to prefer models with such large values of $m_{3/2}$ is that they can automatically solve the moduli problem~\cite{Coughlan:1983ci,deCarlos:1993jw,Randall:1994fr,Acharya:2008bk}, because decays of moduli happen quickly enough for successful BBN and may also produce dark matter with the right relic abundance~\cite{Moroi:1999zb}. In constrast, the most effective known solution to the moduli problem for low-scale SUSY breaking is a late period of inflation~\cite{Randall:1994fr}, but achieving this consistent with all constraints is extremely difficult even when exploiting fields like the saxion that naturally have a nearly flat direction~\cite{deGouvea:1997tn,Kawasaki:1997ah,Fan:2011ua}. Thus, we expect that natural SUSY models with all scales set by $m_{3/2}$ are also the best candidates for reconciling natural SUSY with cosmological constraints.

\subsection{The third generation and natural SUSY}
\label{sec:thirdgen}

So far our discussion has centered on the Higgs sector of the theory. As we have mentioned, another requirement for natural SUSY is an explanation of why the first and second generation superpartners have eluded detection so far, given that the third generation superpartners must be light for naturalness. One resolution is that the third generation superpartners  have soft masses less than those of the first and second generation. Ref.~\cite{Csaki:2012fh} considered a natural model of this sort in which composite states of the magnetic theory were protected at leading order from supersymmetry breaking, so a natural hierarchy of supersymmetry-breaking masses between elementary and composite states is established.

Another possibility would be models in which they are charged differently under flavor symmetries~\cite{Pomarol:1995xc,Dvali:1996rj,Craig:2012di}. However one has to work out the full flavor sector to check consistency of these scenarios.

A more generic possibility suggested by the composite dual scenario is one in which the light states (after supersymmetry breaking) are separated from each other in extra dimensions~\cite{Sundrum:2009gv,Gabella:2007cp,Gherghetta:2011wc,Larsen:2012rq}; the deconstructed analogue~\cite{Craig:2011yk,Auzzi:2011eu,Craig:2012hc,Cohen:2012rm}; in which the third generation is composite~\cite{Delgado:2005fq,Csaki:2011xn,Csaki:2012fh}; or in which the first and second generations are composite (and the composite sector breaks SUSY)~\cite{ArkaniHamed:1997fq,Luty:1998vr,Franco:2009wf,Craig:2009hf,SchaferNameki:2010iz}. Notice that this scenario does not rely on a single extra dimension (though the composite dual interpretation does). This means that any model (such as string type models) in which wavefunctions in a higher-dimensional space determine the quark and lepton masses can in principle fall into this category.

\begin{figure}[h]
\begin{center}
\includegraphics[width = \textwidth]{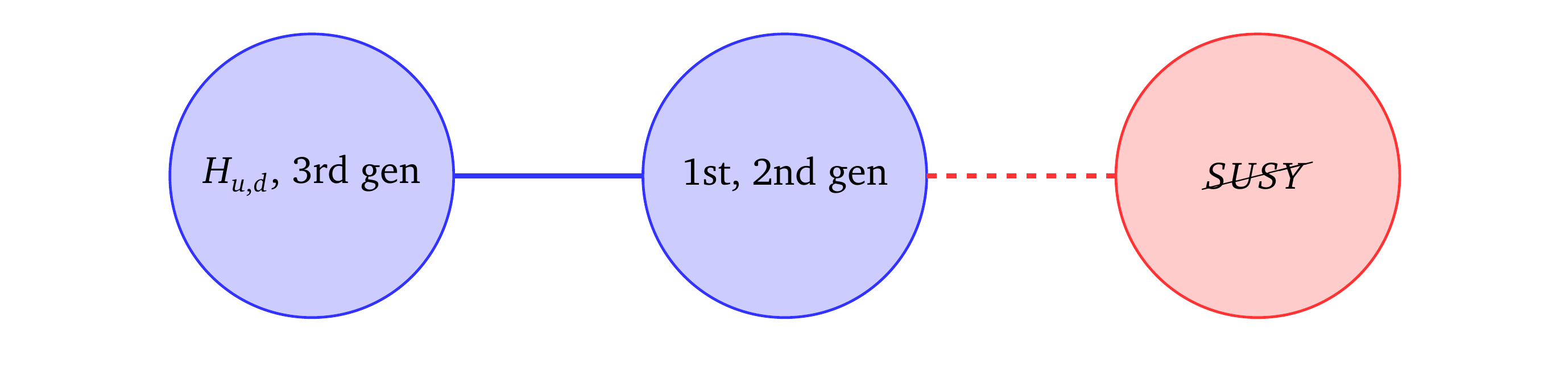}
\end{center}
\caption{A schematic diagram for a viable natural SUSY theory. The third generation quarks are segregated from those of the first and second generation, with SUSY breaking communicating more strongly with the latter, to allow for naturalness without conflicting with direct collider searches. The Higgs fields, having large Yukawa couplings with the third generation, are also expected to be separated from large SUSY breaking effects. In concrete models, this may be interpreted as a moose model with extra $D$-terms, or as a sketch of an extra dimension in which additional light degrees of freedom like a composite singlet $S$ may interact with the Higgs sector.}
\label{fig:schematic}
\end{figure}

We expect that any natural SUSY model with split families consistent with current data will have a schematic structure similar to that of Figure~\ref{fig:schematic}, with a low scale of compositeness that is relevant to the Higgs sector, though the Higgs fields are not necessarily themselves composite. The first and second generation in this setup experience SUSY breaking more strongly than the third generation. Since the Higgses couple more strongly to the third generation, they are also more likely to be insulated from SUSY breaking.

Despite the simplicity of the scenario, two interesting effects tell us that there is a limit to how split we should expect the various MSSM scalar generations to be. Even if we insulate the stops from SUSY breaking and arrange for running only from a very low scale, as in the composite stop model of Ref.~\cite{Csaki:2012fh}, the RG effect of the gluino mass will very quickly bring the stop to a similar scale: $m_{\tilde t}^2 \approx \frac{2}{3\pi^2} g_3^2 M_3^2 \log\frac{\Lambda}{M_3}$, which lifts the stop masses to about 400 GeV if $M_3 \approx 1$ TeV and $\Lambda \approx 10$ TeV. On the other hand, at two loops, the renormalization-group effect of heavy first- and second-generation scalars is to push the third-generation scalars to lower masses~\cite{ArkaniHamed:1997ab,Barbieri:2010pd,Tamarit:2012ie,Tamarit:2012ry,Tamarit:2012yg}. This can increase the possible splitting, but quickly leads to either tachyonic third generation scalars or a need for a large initial soft-mass for the third generation at high RG scales. The latter would reintroduce fine-tuning in the Higgs sector. To avoid these dangerously large two-loop RG effects, following Ref.~\cite{Barbieri:2010pd}, we will take the first- and second-generation squarks to have mass between 5 and 10 TeV (as gluino masses range from about 1 to 2 TeV). In the context of models that add new multiplets charged under SM gauge groups, we will need to revisit the effects of such two-loop terms, which can lead to  important model-building constraints.

In the next section, we will consider a model that can generate the Higgs-sector parameters of a $\lambda$SUSY-like scenario from a single scale $m_{3/2}$. Because our main goal is to understand the Higgs sector, we will not dwell on flavor physics in great detail. However, because LHC data motivates splitting the first two generations of squarks from the third, we will assume their soft masses arise from different sources; in fact, they will be geometrically separated (on different branes) in our construction. Nonuniversal soft terms for different generations potentially cause flavor problems, which have received a great deal of recent attention in the context of natural SUSY. As discussed in Ref.~\cite{Csaki:2012fh}, the range of first- and second-generation squark masses between 5 and 10 TeV, keeping the third generation below 1 TeV, can be safe from most flavor constraints as studied in~\cite{Giudice:2008uk} if the first and second generation squarks are degenerate. These masses also avoid dangerous two-loop RG effects~\cite{Barbieri:2010pd}. This is the sweet spot in which our theory should live. Recently, it has been pointed out that in natural SUSY the right-handed sbottom must also be rather heavy, $m_{{\tilde b}_R} \simgt 4$ TeV, to avoid dangerous CP-violating contributions to $K-{\bar K}$ mixing assuming order-one phases~\cite{Brust:2011tb}. This will either imply that our right-handed sbottom must have a wavefunction localized near the source of SUSY breaking (in contrast to the other third generation sfermions) or that CP-violating phases in the squark mass matrix are small. Minimizing CP-violating phases is also preferable from the point of view of EDM constraints. In general, when separating the first and second generation squarks from the third generation, it is useful to impose some flavor symmetries on the light generations~\cite{Barbieri:2010ar,Barbieri:2011vn,Barbieri:2011ci}. Because we are considering large values of $m_{3/2}$, one might expect generic Planck-suppressed operators to spoil such flavor structures. Arbitrary $M_P$-suppressed operators could split the first and second generations and conflict with strong constraints in the kaon sector. However, it turns out that effective supergravity theories arising from string compactifications often have a non-generic structure related to the special form of moduli couplings, which in some cases can protect approximate flavor symmetries from dangerous effects of SUSY breaking~\cite{Conlon:2007dw} (see also~\cite{Kadota:2011cr}). Hence, we will proceed under the assumption that the light generations will have approximately degenerate soft scalar masses. Alternatively, we can assume that {\em all} the generations have degenerate soft masses, which are light as required by naturalness, but that they have gone undetected due to a mechanism that hides them from collider searches, like $R$-parity violation. After discussing the Higgs sector of our model in detail, we will revisit the flavor issues with some brief remarks in Section~\ref{subsec:flavorconstraints}.

\section{Single-scale NMSSM}
\label{sec:singlescale}

\begin{figure}[h]
\begin{center}
\includegraphics[width = \textwidth]{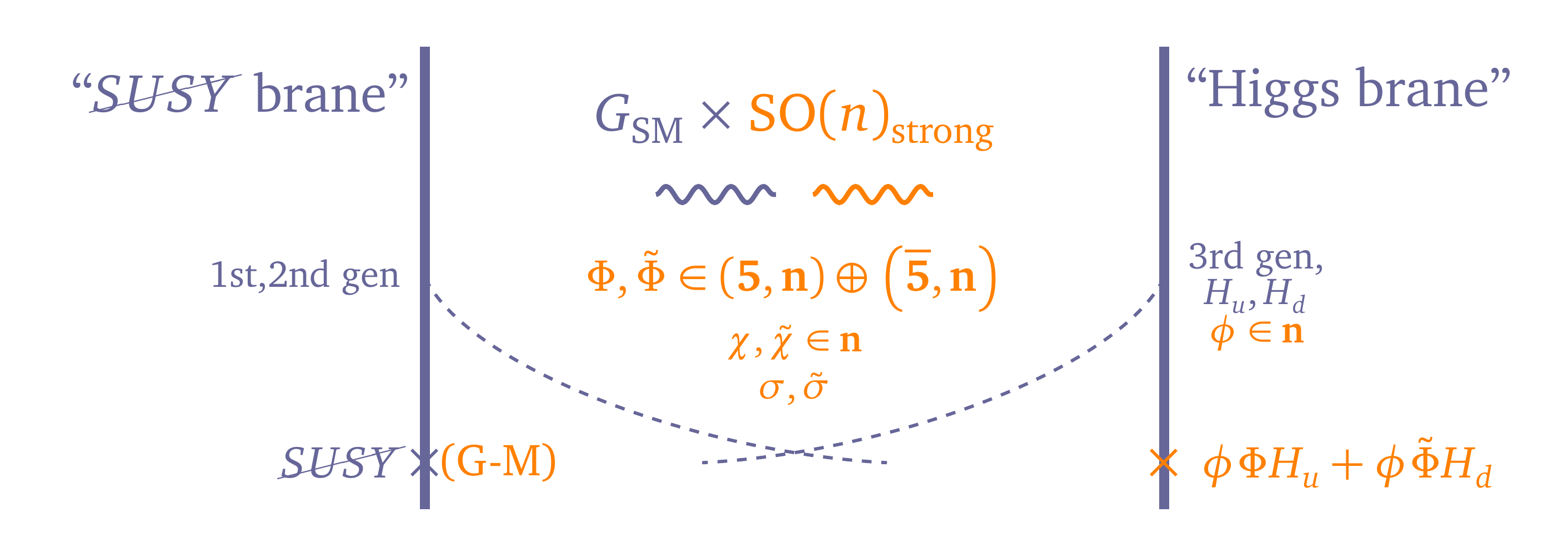}
\end{center}
\caption{Geometry of our composite model. Ingredients beyond the MSSM are color-coded orange. Gauge fields for both the Standard Model and the new strong gauge group propagate in the bulk, as do fields in the bifundamental. Standard Model fields are localized near branes, with the first and second generation localized near SUSY breaking (the $\cancel{SUSY}$ brane, on the left) and the third generation and the Higgses separated from SUSY breaking (near the Higgs brane, on the right). The dashed lines illustrate possible wavefunctions, peaked near a brane and trailing off into the bulk. SO($n$) matter is also localized on the Higgs brane, interacting with the Higgs and bulk bifundamental modes. The bifundamentals, along with other bulk fields, have Giudice-Masiero masses (``G-M'') through their overlap with the $\cancel{SUSY}$ brane.}
\label{fig:geometry}
\end{figure}

We would like to consider a $\lambda$SUSY-like scenario, i.e. a version of the NMSSM that has a relatively large value of $\lambda$, as naturalness considerations prefer~\cite{Hall:2011aa}. Unlike the original $\lambda$SUSY model, we will aim to have a linear superpotential term in $S$ and a tadpole, like Refs.~\cite{Harnik:2003rs,Csaki:2012fh}, so that EWSB is easily achieved.   We ask what is the minimal model which leads to a simpler possibility, namely to assume that {\em only} the singlet $S$, among all NMSSM degrees of freedom, is composite. This approach was taken in Ref.~\cite{Chang:2004db}, which built a model including a number of mass scales put in by hand. Our goal here is to show that a very similar model can work, with all mass scales arising from $m_{3/2}$ a loop factor above the soft mass scale of MSSM fields. To explain why certain fields in the theory obtain masses at $m_{3/2}$, while others see the scale $m_{3/2}$ only indirectly through loops, we imagine the geometric picture illustrated in Figure~\ref{fig:geometry}. Note that the model of Ref.~\cite{Csaki:2012fh}, although four-dimensional, is dual to a similar picture in the conformal regime. (Here we aren't necessarily taking the space to be warped, however.)

We will begin by explaining the model at a big-picture level, leaving a discussion of subtle but important details to the following subsections. Our discussion will approximately follow~\cite{Chang:2004db}, with a few differences . We begin with a superpotential
\beq
W = \lambda_1 \phi \Phi H_u + \lambda_2 \phi {\tilde \Phi} H_d + y \left( \phi \chi \sigma + \phi {\tilde \chi} {\tilde \sigma}\right),
\label{eq:oursuperpotential}
\eeq
where the $\phi$ fields are SM singlets but charged under a new SO($n$) gauge group, and the $\Phi$ fields are bifundamentals of SU(2) and SO($n$) (with appropriate hypercharge). We assume SO($n$), among other reasons, so that eventually we will have only composite mesons to deal with, rather than unwanted composite baryons that would be massless without additional structure in an SU($n$) theory. We will integrate out the (massive) $\Phi$ fields, after which SO($n$) will confine and turn $\phi^2$ into our singlet $S$. The fields $\chi$ and $\sigma$ play a role in generating tadpoles and linear terms for $S$, but let us first discuss the $\lambda S H_u H_d$ term.

We want the mass scale at which we integrate out $\Phi$ to not be set by hand (as it was in~\cite{Chang:2004db}) but to come from $m_{3/2}$, as in the Giudice-Masiero mechanism~\cite{Giudice:1988yz}:
\beq
{\cal L} \supset \int d^4\theta \frac{X^\dagger}{M_{\rm Pl}} \Phi {\tilde \Phi} + c \frac{X^\dagger X}{M_{\rm Pl}^2} \Phi {\tilde \Phi}  \to \int d^2\theta M \Phi {\tilde \Phi} + c.c.
\label{eq:GMtosup}
\eeq
where $X$ has an $F$-term of order $m_{3/2} M_{\rm Pl}$. The first term generates a ``$\mu$-term'' for the $\Phi$ fields while the second generates a $B$ term (here, and throughout this section, terms labeled $B$ will be dimension two. What we denote as $B$ is often denoted $B\mu$ in the literature). The arrow in Eq.~\ref{eq:GMtosup} indicates that we can think of this an effective superpotential mass $M = M_0 + \theta^2 B_0$ that encodes the effect of both the mass and the $B$-term. Then when we integrate out the $\Phi$ fields, we obtain an effective superpotential:
\beq
W_{\rm eff} = \frac{\lambda_1 \lambda_2}{M} \phi^2 H_u H_d.
\eeq
Now, SO($n$) confines, which leads to a replacement (by NDA~\cite{Luty:1997fk,Cohen:1997rt}):
\beq
\phi^2 \to \frac{\Lambda \sqrt{n}}{4\pi} S,
\eeq
with $\Lambda$ the (holomorphic) dynamical scale of the theory. In particular, this means that confinement gives us
\beqs
W_{\rm eff} & = & \frac{\lambda_1 \lambda_2}{M} \phi^2 H_u H_d \\
& = & \frac{\sqrt{n}}{4\pi} \frac{\lambda_1 \lambda_2 \Lambda}{M} S H_u H_d \\
& = & \frac{\sqrt{n}}{4\pi} \lambda_1 \lambda_2 e^{-\frac{8\pi^2}{bg_*^2}} S H_u H_d.
\eeqs
We assume the gauge coupling $g_*$ is relatively large, which suggests the theory should be at or near a conformal fixed point at energies about the scale $M$.

We would like to generate a theory that has an effective superpotential of the form
\beq
W_{\rm eff} = \lambda S H_u H_d - f S + \ldots,
\eeq
as well as a tadpole for $S$, as such models offer one of the cleanest realizations of electroweak symmetry breaking~\cite{Harnik:2003rs,Csaki:2012fh}. In Ref.~\cite{Chang:2004db}, this was achieved by simply turning on a mass for fundamentals of the electric $SU(n)$ theory, which after confinement become a linear term in $S$. We could do this, but it introduces a mass scale by hand, which we are trying to avoid. Instead, we produce a similar effect from dynamics. Because $S$ is a $\phi^2$ composite, we can try to build terms that generate a $\phi^2$ term when fields are integrated out, again relying on Giudice-Masiero. In particular, we assume new bulk fields $\chi, {\tilde \chi}$ that are SO($n$) fundamentals and $\sigma, {\tilde \sigma}$ that are SO($n$) singlets, with a superpotential
\beq
W = y \left( \phi \chi \sigma + \phi {\tilde \chi} {\tilde \sigma}\right)
\eeq
and Giudice-Masiero masses for $\chi {\tilde \chi} $ and $\sigma {\tilde \sigma}$. The Giudice-Masiero $B$-term masses break supersymmetry, and a loop calculation shows that this generates effective $\phi^2$ terms that become $S$ terms after confinement:
\beq
\int d^4x \left[B_{\rm eff} \phi^2 + \int d^2\theta m_{\rm eff}^2 \phi^2 + c.c. \right]  \to \int d^4 x \left[ T S + \int d^2\theta f S + c.c.\right],
\eeq
where $fS$ is an effective linear superpotential term with
\beq
f \sim \frac{y^2}{32\pi^2} \frac{B}{\mu}\Lambda
\eeq
and $TS$ is an effective SUSY-breaking tadpole term with
\beq
T \sim \frac{y^2}{32\pi^2} \left(\frac{B}{\mu}\right)^2 \Lambda.
\eeq
The tadpole drives $S$ to get a VEV, producing a sizable $\mu$ term much more easily than in the ${\mathbb Z}/3$-symmetric NMSSM; the $f$-term provides a VEV for $H_u H_d \sim f$, favoring $\tan\beta \approx 1$, the regime in which the  $\lambda$ quartic is most effective at raising the Higgs mass.

From this sketch of the model, it would appear that it works beautifully. However, there are a few crucial subtleties, which we will spend the next subsections exploring. One is that in order for fields like $\Phi$, $\chi$, and $\sigma$ to effectively communicate SUSY breaking from the ${\cancel{SUSY}}$ brane to the Higgs brane where the fields in our low-energy effective theory live, they must have relatively flat profiles in the extra dimension. Otherwise, these terms would be exponentially suppressed. This means they have a bulk mass that is not too large in units of $1/L$, the radius of the extra dimension. The other point is that in order for $\Lambda/M$ not to be so small that $\lambda$ is ineffective at generating a heavy Higgs, we need $g_*$ to be reasonably large, which leads to strong renormalization group effects in the conformal window above $M$. This tends to enhance many of the terms in our low-energy effective theory, while the wavefunction overlaps threaten to suppress them. In the end, we will balance these terms, but it implies a restriction on the parameter space that shows the model is not quite as simple as it at first appears.

\subsection{Canceling dangerous $A$-terms}
\label{sec:dangerA}

One danger in this theory is that the Giudice-Masiero mass is $M = M_0 + \theta^2 B_0$. When we integrate out fields to produce a higher dimension operator in $W_{\rm eff}$ involving $M$, it also generally has a $\theta^2$ component, so $\int d^2\theta~\lambda S H_u H_d$ can be accompanied by the trilinear scalar term $A_\lambda S H_u H_d$. Because $B_0/M_0$ is of order $m_{3/2} \sim 30$ TeV, this $A_\lambda$ term could be so large that it completely overwhelms the other weak-scale SUSY breaking terms we would like to have in the visible sector. Luckily, it turns out that $A_\lambda$ is suppressed. As discussed above, when we integrate out the $\Phi$ fields, we obtain:
\beq
W_{\rm eff} = \frac{\lambda_1 \lambda_2}{M} \phi^2 H_u H_d = \frac{\sqrt{n}}{4\pi} \frac{\lambda_1 \lambda_2 \Lambda}{M} S H_u H_d,
\eeq
and if $M$ is the only mass scale taking us out of the conformal window we have $\Lambda = e^{-\frac{8\pi^2}{bg_*^2}} M$, so if we write $\Lambda = \Lambda_0 + F_\Lambda \theta^2$, we have $\Lambda/M = \left(\Lambda_0 + \theta^2 F_\Lambda\right)/\left(M_0 + \theta^2 B_0\right) = \Lambda_0/M_0$, with the $\theta^2$ pieces canceling between numerator and denominator.

On the other hand, if we integrate out some SO($n$) charged fields at one mass scale, and some at another, we need to be a little more careful. If we first integrate out some fields at $M_1$ such that the beta function coefficient becomes $b_1$, and then integrate out more fields at $M_2 < M_1$ such that the beta function coefficient becomes $b_2$, then we have:
\beq
\Lambda = M_1 \left(\frac{M_2}{M_1}\right)^{\frac{b_1+b_2}{b_2}} e^{-\frac{8\pi^2}{b_2 g_*^2}},
\eeq
and
\beq
\frac{F_\Lambda}{\Lambda} = -\frac{b_1}{b_2} \frac{B_1}{M_1} + \frac{b_1 + b_2}{b_2} \frac{B_2}{M_2}.
\label{eq:Flambda}
\eeq
Hence, in the limit when $B/M$ is the same for all the fields we integrate out, there is no induced $A$-term. Thus, we will assume a symmetry among $\chi,{\tilde \chi}$ and $\Phi, {\tilde \Phi}$, such that they have the same ratio $B/M$ in their Giudice-Masiero terms. The symmetry is broken by the fact that $\Phi, {\tilde \Phi}$ transform under Standard Model gauge groups while $\chi, {\tilde \chi}$ do not, but this induces only small perturbative corrections, allowing $A_\lambda$ to remain a loop factor below $m_{3/2}$.

\subsection{SO($n$) dynamics}
\label{subsec:SOndynamics}

As we have already noted, both wavefunction overlap factors in the extra dimension and renormalization group effects from strong dynamics play an important role in our effective theory. Higher-dimensional wavefunction overlaps act to suppress couplings, since fields are attenuated as they propagate through the extra dimension. Meanwhile, strong dynamics enhances couplings as the theory flows toward lower energies. We'll introduce small numbers $\psi$ from extra-dimensional wavefunction suppression and $\epsilon$ related to RG effects. Their appearance in various terms in the theory is summarized in Table~\ref{tab:factors}. One can see that, although they complicate the parametric simplicity of relying on one scale, we can always play small factors of $\psi$ or small Yukawas off against $\epsilon^{-1}$ RG enhancements, so the theory is viable. Our goal now is to explain these factors in more detail. We'll begin with a closer look at the strong SO($n$) dynamics.

\begin{table}[ht]
\centering
\large
\begin{tabular}{l|lll}
Quantity & Couplings & Wavefunction & Conformal \\
& & overlap & dynamics \\
\hline
$M_\Phi$ & $c_\mu$ & $\left(\psi^{\cancel{SUSY}}_\Phi\right)^2$ & $\epsilon^{-1}$ \\
$B_\Phi$ & $c_B$ & $\left(\psi^{\cancel{SUSY}}_\Phi\right)^2$ & $\epsilon^{-1}$ \\
$\Lambda$ & $e^{-\frac{8\pi^2}{b g^2}}M$ & $\left(\psi^{\cancel{SUSY}}_\Phi\right)^2$ & $\epsilon^{-1}$ \\
$\lambda$ & $\lambda^0_1 \lambda^0_2 \frac{\Lambda}{M_\Phi}$ & $\left(\psi^{\rm Higgs}_\Phi\right)^2$ & $\epsilon^{-2}$  \\
$A_\lambda$ & $\lambda^0_1 \lambda^0_2 \frac{\Lambda \delta_A B}{M_\Phi^2}$ & $\left(\psi^{\rm Higgs}_\Phi\right)^2$ & $\epsilon^{-2}$ \\
$f$ & $\frac{y_0^2}{32\pi^2}\frac{\Lambda B}{M}$ & $\left(\psi^{\rm Higgs}_\chi \psi^{\rm Higgs}_\sigma\right)^2$ & $\epsilon^{-3}$ \\
$T$ & $\frac{y_0^2}{32\pi^2}\frac{\Lambda \delta_T B^2}{M^2}$ & $\left(\psi^{\rm Higgs}_\chi \psi^{\rm Higgs}_\sigma\right)^2$ & $\epsilon^{-3}$
\end{tabular}
\caption{Numerical factors affecting the scaling of quantities in the low-energy effective theory. As explained near Eqns.~\ref{eq:Flambda} and~\ref{eq:tadpole}, the terms $A_\lambda$ and $T$ vanish in the limit that all fields have equal $B/\mu$; this is accounted for by the factors $\delta_A$ and $\delta_T$. We have omitted the $\sqrt{n}/(4\pi)$ factor accompanying each $\Lambda$ from NDA. We have factored wavefunction overlaps out of couplings, so that e.g. the value $\lambda_1$ in the low-energy effective theory is $\lambda_1^0 \psi^{\rm Higgs}_\Phi$ and $y = y_0 \psi^{\rm Higgs}_\chi \psi^{\rm Higgs}_\sigma$.}
\label{tab:factors}
\end{table}

After we integrate out some massive fields, we would like to have one or more nearly composite SM gauge singlet states, one of which, $S$, plays the role of the NMSSM singlet. We will make use of the fact that an SO($n$) gauge theory with $n-4$ flavors has vacua in which composite mesons, $M^{ij} = Q^i \cdot Q^j$, are free (with no superpotential)~\cite{Intriligator:1995id}. These mesons form a symmetric matrix constructed from $n-4$ real fields, and so there are $\frac{1}{2} (n-4)(n-3)$ of them.

At high energies, we have a larger number of flavors; we denote this number as $n_f$. We assume that the theory is in the conformal window at these energies, which implies $\frac{3}{2}\left(n-2\right)\leq n_f \leq 3\left(n-2\right)$. In Section~\ref{subsec:SOnEWSB}, we will introduce two flavors of SO($n$) that play a role in generating terms in the potential for the singlet. Thus, we will consider two scenarios:
\begin{itemize}
\item {\bf Minimal model}: We need one light flavor to generate our singlet $S$, four flavors (two doublets) to couple to $H_u$ and $H_d$, and two more flavors to play a role in generating the $S$ tadpole. This leads us to consider SO(5) with 7 flavors (in the middle of the conformal window).
\item {\bf Unified model}: Rather than adding two doublets to couple $H_u$ and $H_d$, we aim to keep the successful MSSM gauge coupling unification, so we add fields in the fundamental of SO($n$) that are in a ${\bf 5}$ and ${\bf \overline{5}}$ of SU(5)$_{\rm GUT}$. This brings the total necessary flavor count to 13, which is too large to fall into the SO(5) conformal window. So we need a larger group; for instance, we can consider SO(11) with 19 flavors, of which we integrate out 12, leaving a low energy theory with 7 flavors. This gives rise to 28 mesons, of which only one is needed to play the role of our singlet.
\end{itemize}
Clearly, in this setting, keeping gauge coupling unification comes at a steep cost, as we must find a way to integrate out the extra unwanted matter so that it is no longer relativistic at the time of BBN.

We  work out the one-loop estimate of the fixed point coupling $g_*$ and the confinement scale after integrating out the heavy matter fields in Appendix~\ref{app:SOnbasics}. The results for the two models we have discussed are summarized in Table~\ref{tab:SOnproperties}.
\begin{table}[ht]
\centering
\large
\begin{tabular}{l|ll|lll}
Model & $n$ & $n_f$ & $\gamma$ & $\alpha_* (n-1)$ & $\Lambda/M$ \\
\hline
Minimal model & 5 & 7 & $-\frac{2}{7}$ & 1.8 & 0.4\\
Unified model & 11 & 19 & $-\frac{8}{19}$ & 2.6 &  0.5\\
\end{tabular}
\caption{Properties of the conformal and confining phases of the two model we consider. $\alpha_* (n-1) = \frac{g_*^2}{4\pi} (n-1)$, with $g_*$ the one-loop value of the fixed-point coupling, is taken as a rough estimate of how strongly coupled the theory is.}
\label{tab:SOnproperties}
\end{table}

The next question we should address is the effect of the anomalous dimension $\gamma$ on the dynamics of the theory. The superpotential operators, like $\lambda_1 \phi \Phi H_u$, are not renormalized in the holomorphic basis. Similarly, although the Giudice-Masiero operators like $X^\dagger X \Phi {\tilde \Phi}$ are potentially subject to hidden-sector renormalization if $X$ interacts strongly with other fields, they are insensitive to the anomalous dimensions of $\Phi$ and ${\tilde \Phi}$ since they are holomorphic in these operators~\cite{Roy:2007nz,Murayama:2007ge}. Nonetheless, $\Phi$ has an anomalous dimension, and it does have an effect on the theory.

Let's begin with the spectrum of the $\Phi$ multiplet itself. In addition to the Giudice-Masiero terms, it will also in general contain a soft mass, and all of these interactions can be suppressed by a factor $\left(\psi^{\cancel{SUSY}}_\Phi\right)^2$ from the wavefunction overlap of $\Phi$ with the $\cancel{SUSY}$ brane:
\beq
{\cal L}_\phi = \int d^4\theta \left(1 +  \left(\psi^{\cancel{SUSY}}_\Phi\right)^2 c_s \frac{X^\dagger X}{M_{\rm Pl}^2}\right) Z \left(\Phi^\dagger \Phi + {\tilde \Phi}^\dagger {\tilde \Phi}\right) + \left(\psi^{\cancel{SUSY}}_\Phi\right)^2 \left(c_\mu \frac{X^\dagger}{M_{\rm Pl}} \Phi {\tilde \Phi} + c_B \frac{X^\dagger X}{M_{\rm Pl}^2} \Phi {\tilde \Phi} + h.c. \right).
\eeq
Strong dynamics will renormalize the factor $Z$ and only the factor $Z$, which will scale as $\left(\frac{\mu}{\Lambda_{*}}\right)^{-\gamma}$ where $\gamma$ is the anomalous dimension as reported in Table~\ref{tab:SOnproperties}. In terms of the {\em canonically normalized fields}, then, and assuming $F_X \sim m_{3/2} M_{\rm Pl}$, we have:
\beqs
m_{\rm soft}^2 & \sim & \left(\psi^{\cancel{SUSY}}_\Phi\right)^2 c_s m_{3/2}^2\\
\mu & \sim & \left(\psi^{\cancel{SUSY}}_\Phi\right)^2 c_\mu m_{3/2} \left(\frac{\Lambda_{*}}{\mu}\right)^{-\gamma} \\
B & \sim & \left(\psi^{\cancel{SUSY}}_\Phi\right)^2 c_B m_{3/2}^2 \left(\frac{\Lambda_{*}}{\mu}\right)^{-\gamma}
\eeqs
Here $\Lambda_{*}$ should be interpreted as the scale at which the interacting SO($n$) theory approaches its conformal fixed point. As a general rule, we require that $B < \mu^2$, as we risk tachyonic scalars otherwise. Giudice-Masiero naturally gives us $\mu^2 \sim B$, but here both $\mu$ and $B$ are enhanced by the same factor $\left(\psi^{\cancel{SUSY}}_\Phi\right)^2 \left(\Lambda_*/\mu\right)^{-\gamma}$. Thus, we require that this factor is {\em larger than one}, so that the ratio $\mu^2/B$ only increases. Because $\gamma < 0$, the RG effect gives a potentially large enhancement of the $\mu$ term, an enhancement of $B$ smaller than that of $\mu^2$, and no enhancement of the soft mass. Since the soft mass could potentially drive the stops tachyonic at two loops, this is a welcome development.

\begin{figure}[!h]
\includegraphics[width = \textwidth]{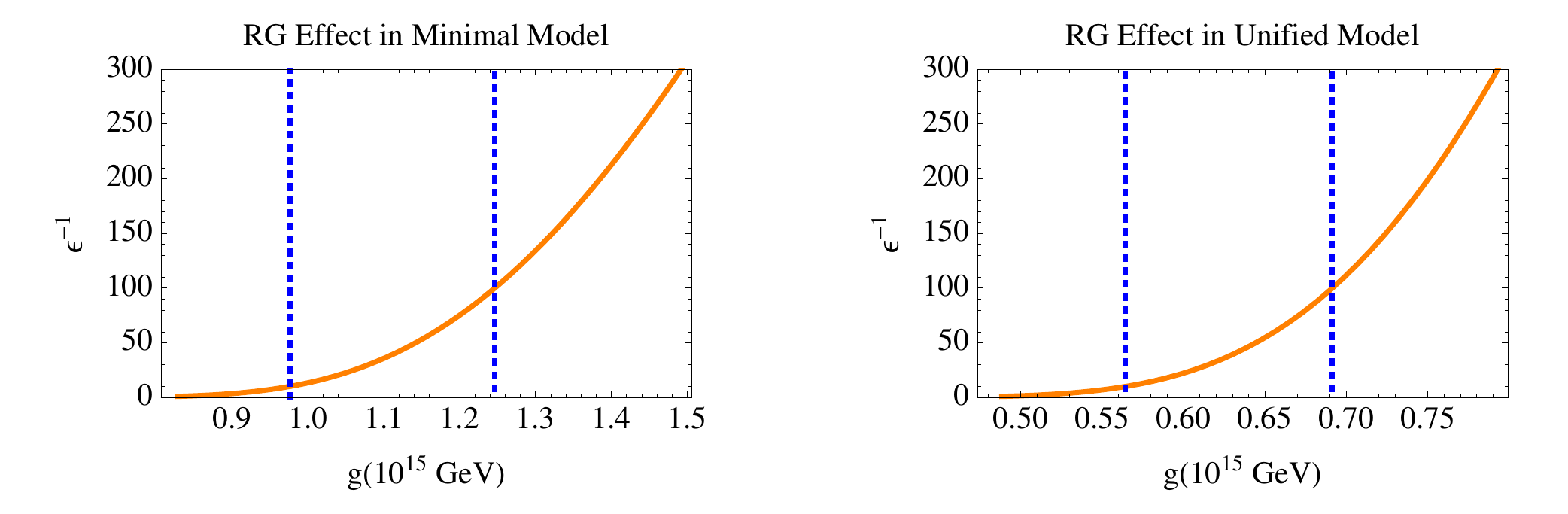}
\caption{The renormalization group factor $\epsilon^{-1} = \left(\frac{\Lambda_{*}}{\mu}\right)^{-\gamma}$. Here we assume the SO($n$) gauge coupling begins at the value $g(10^{15}~{\rm GeV})$ on the horizontal axis, reaches its conformal value at $\Lambda_{*}$, and departs from the conformal window at $\mu = 10$ TeV. We ignore running effects above $\Lambda_{*}$. The dotted vertical lines bound the values of $g$ at the high scale for which the enhancement factor is between 10 and 100.}\label{fig:RGeffect}
\end{figure}

We will keep track of RG enhancements by counting powers of a small parameter $\epsilon$:
\beq
\epsilon \equiv \left(\frac{\Lambda_{*}}{\mu}\right)^\gamma.
\eeq
This parameter is a sensitive function of the input value for the gauge coupling at high scale. We can estimate it assuming that the gauge coupling begins at some value weaker than its conformal fixed point value $g_*$ at a high scale, e.g. $10^{15}~{\rm GeV}$, runs to $g_*$ at $\Lambda_{*}$, and remains there until the scale $\mu \approx 10$ TeV at which fields are integrated out. The result of a simple one-loop estimate of $\epsilon^{-1}$ is displayed in Figure~\ref{fig:RGeffect}. If $g$ is too small, the gauge coupling never reaches the conformal value. For higher values of $g$, the enhancement factor increases rapidly. Nonetheless, there is a range of reasonable values of $g$ for which the enhancement is a factor between 10 and 100, marked by dotted lines on the plots. We will focus on values in this range. In order that the wavefunction overlap factors do not spoil the relation $B < \mu^2$, we require $\left(\psi^{\cancel{SUSY}}\right)^2 \simgt \epsilon$. In particular, we can consider smaller values of $g$, so that the theory barely reaches the conformal window and $\epsilon \approx 1$, requiring $\psi^{\cancel{SUSY}} \approx 1$ as well.

The next point is that $W_{\rm eff} = \frac{\lambda_1 \lambda_2}{M} \phi^2 H_u H_d$ is true in the holomorphic basis, with $M$ the effective mass term for $\Phi {\tilde \Phi}$. However, we should be careful in canonically normalizing $S$ given the wavefunction renormalization of $\phi$, as well as of the field $\Phi$ that we have integrated out. Wavefunction renormalization enhances $\lambda_1$ and $\lambda_2$ by powers of $\epsilon^{-2}$; $\Lambda$ and $M$ are both enhanced by $\epsilon^{-1}$, which cancel. Similarly, the wavefunction overlap $\left(\psi^{\cancel{SUSY}}_\Phi\right)^2$ appears in $\Lambda$ and $M$ and cancels, whereas a wavefunction overlap factor $\phi^{\rm Higgs}_\Phi$ of $\Phi$ with the brane where $H_u,~H_d$, and $\phi$ are localized appears once in both $\lambda_1$ and $\lambda_2$. The net result is that $\lambda$ scales as $\left(\psi^{\rm Higgs}_\Phi\right)^2 \epsilon^{-2}$. The $\epsilon^{-2}$ factor is rather large, suggesting that we either assume small Yukawa couplings $\lambda_{1,2}$ on the brane, or a small wavefunction overlap $\left(\psi^{\rm Higgs}_\Phi\right)^2$.

\subsection{SO($n$) model: EWSB}
\label{subsec:SOnEWSB}

We have seen how the $\lambda S H_u H_d$ term can be generated. Now, we also want to generate a linear term $f S$ in the superpotential, in order to have a simple model for electroweak symmetry breaking. We can do that by adding more bulk matter that feels SUSY breaking and communicates with $\phi$ in a different way. For instance: suppose we add new fields $\chi, {\tilde \chi}$ which are SO($n$) fundamentals. They are also charged under some other symmetry, say a U(1) or a discrete symmetry, which enforces the couplings:
\beq
y\left(\phi \chi \sigma + \phi {\tilde \chi} {\tilde \sigma}\right),
\eeq
where $\sigma, {\tilde \sigma}$ are fields not charged under SO($n$). Now, we also add Giudice-Masiero masses for $\chi {\tilde \chi}$ and $\sigma {\tilde \sigma}$. We imagine that the fields $\chi,~{\tilde \chi}$ and $\Phi,~{\tilde \Phi}$ have the same Giudice-Masiero terms and the same bulk mass, in order to ensure their $B/\mu$ ratio is equal, as discussed near Eq.~\ref{eq:Flambda}. Then a loop will generate a $\phi \phi$ mass which becomes the desired linear term for $S$, as well as a tadpole term for $S$ that leads to a VEV. The details of the loop calculation are presented in Appendix~\ref{app:loops}. The resulting parametric scaling is:
\beqs
f & \approx & \frac{\sqrt{n}}{4\pi} \frac{y_0^2}{32\pi^2} \frac{B \Lambda}{\mu} \left(\psi^{\rm Higgs}_\chi \psi^{\rm Higgs}_\sigma\right)^2 \epsilon^{-3}, \\
T & \approx & f \frac{\delta B}{\mu},
\eeqs
where $B,~\mu$ are typical Giudice-Masiero terms of the theory (before taking into account wavefunction renormalization and overlap factors) and $\delta B$ indicates that $T$ vanishes in the limit where $B/\mu$ is identical for all fields.

We are now in a position to try to put everything together. In terms of the low-energy effective theory, one set of numbers that gives a good solution in the tree-level potential: $\lambda = 1.1$, $f = \left(100~{\rm GeV}\right)^2$, $T = 1.8 \times 10^6~{\rm GeV}^3$, $A_\lambda = 200~{\rm GeV}$, $m_{H_u}^2 = -\left(70~{\rm GeV}\right)^2$, $m_{H_d}^2 = \left(120~{\rm GeV}\right)^2$, $m_S^2 = \left(100~{\rm GeV}\right)^2$. This leads to $\tan \beta = 1.7$, a 121 GeV mostly-up-type Higgs, and Higgses at 214 and 252 GeV that are mixtures of mostly $S$ and $H_d$. The effective $\mu$-term $\lambda\left<S\right> = -148$ GeV.

The simplest regime to study the theory  would be that in which the bulk fields $\Phi, \chi, \sigma$ have zero bulk mass and hence flat wavefunction profiles, so all the $\psi$ factors are near 1, and the gauge coupling has just reached its conformal value near $m_{3/2}$, so $\epsilon \approx 1$. But this is clearly just compounding the ``coincidence of scales" problem we aimed to avoid. A more reasonable choice is that the gauge coupling remains near its conformal value over some regime, with $\epsilon$ somewhat small, and couplings and wavefunctions adjusted to partially compensate. To attain these numbers, we can take, following Table~\ref{tab:SOnproperties}, $\Lambda/M = 0.4$ and follow the scalings in Table~\ref{tab:factors}. We will fix $m_{3/2} = 25$ TeV, $c_\mu \left(\psi^{\cancel{SUSY}}_\Phi\right)^2 = c_B \left(\psi^{\cancel{SUSY}}_\Phi\right)^2 = 0.1$, $\epsilon^{-1} = 20$. Then: $M_\Phi = 50$ TeV, $B_\Phi = \left(35~{\rm TeV}\right)^2$, and $\Lambda = 20$ TeV. We will take the $\psi$ parameters to be equal, $\psi^{\rm Higgs}_{\sigma,\chi} = e^{-3} \approx 0.05$. Then if we choose $\lambda^0_1 = \lambda^0_2 = 1. 7$ and $y_0 = 0.36$, all the numbers work out to give the parameters discussed in the previous paragraph. We require the parameters $\delta_T = \delta_A \approx 10^{-2}$, which is reasonable if initially all the fields have the same Giudice-Masiero terms and they are split by radiative effects. Note that we can interpret the $\psi$ factors as the square of a wavefunction in the fifth dimension, and so $\psi \sim e^{-3}$ roughly means:
\beq
M_{\rm bulk} L \approx 3.
\eeq
This suggests that the cutoff scale in the bulk is not far above the compactification scale $L^{-1}$.

One further point remains: in the minimal model, $S$ was the only meson in the low-energy effective theory. However, the unified model had more mesons; in the case we considered, there were 7 flavors of SO($n$) fundamentals in the low-energy effective theory, and thus 28 mesons. Unification is not our main goal in this paper, so we will not go into a detailed discussion of the physics of the remaining mesons here. We have made some brief remarks about how to prevent the remaining mesons from posing a problem at the time of BBN in Appendix~\ref{app:removal}.

\subsection{Flavor constraints}
\label{subsec:flavorconstraints}

As we noted in Sec.~\ref{sec:thirdgen}, models of natural SUSY potentially have problems with flavor physics. A full analysis of flavor physics in our model is beyond the scope of this paper. Our main goal has been to illustrate the possibility of generating a $\lambda$SUSY-like Higgs sector (including an $S$ tadpole and $F$-term for simple EWSB) in which the only energy scale that we input is $m_{3/2}$. As we have seen, this is possible, with some complications from wavefunction overlap and strong dynamics factors. Due to the extra dimensional scenario, we can locate the first and second generations and the third generation in different places and explain why the soft masses for the stops can be much smaller than for the other squarks, as is needed for natural SUSY. However, it is clear that separating the generations in this way can potentially lead to flavor problems, and since we have a high scale of SUSY breaking we must offer some explanation of why there are not generic flavor-violating Planck-suppressed operators. Alternatively, we could imagine that all the generations are light but that they are hidden from collider searches.

Here we will briefly outline some of the issues. First, we note that only the singlet in our model is composite, so unlike in models with composite stops at a low scale~\cite{Csaki:2011xn,Csaki:2012fh} there is no possibility of extra flavor violation from the exchange of flavored composites. Because our squarks propagate in an extra dimension, there are Kaluza-Klein modes with flavor, but these have mass at the compactification scale $1/L$ which can be very large. (A lower choice of $1/L$ could possibly ameliorate some tuning effects that are made worse by having many decades of RG running, but in any case we will choose $1/L$ high enough that generic operators suppressed by this scale are not in tension with flavor constraints.) Thus, the flavor problems that we have are shared with most other models of natural SUSY.

The structure of the soft terms required in theories of split families has been discussed in Ref.~\cite{Giudice:2008uk}. The flavor violation is typically characterized by matrices $\delta_{LL}$ and $\delta_{RR}$ (because the chirality-violating matrices $\delta_{LR,RL}$ can be small). When the $LL$ and $RR$ terms are comparable, bounds from flavor physics are quite strong. For instance, if $\left(\delta_{LL}^d\right)_{12} \approx \left(\delta_{RR}^d\right)_{12} \approx 0.22$ (parametrically set by the Cabibbo angle), the bound from $\Delta m_K$ requires the first and second generation squarks to have masses of order 60 TeV or more~\cite{Kersten:2012ed}.  This can be dangerous, since such large masses lead to large negative contributions to the stop soft mass in the RGE at two loops~\cite{ArkaniHamed:1997ab}. Similar constraints involving mixing with the third generation require sbottoms are heavier than about 3 TeV, if the left- and right-handed sbottoms have comparable masses~\cite{Mescia:2012fg} (see also Ref.~\cite{Brust:2011tb}). If the left-handed sbottom were this heavy, it would directly threaten naturalness, although the right-handed sbottom can have a large mass consistent with naturalness.

The bounds are weaker if $\delta_{LL} \neq 0$ but $\delta_{RR} \approx 0$. In this case, the bound from $\Delta m_K$ merely requires the first and second generation squark masses to be larger than about 6 TeV~\cite{Kersten:2012ed} and the bound on sbottom masses disappears~\cite{Mescia:2012fg}. This is what we require for our model to make sense in the context of split squark generations. One way this could arise is through the U(2)$^3$ framework of Refs.~\cite{Barbieri:2010ar,Barbieri:2011vn,Barbieri:2011ci}, which builds on older work on horizontal (flavor) symmetries in supersymmetric theories~\cite{Pomarol:1995xc,Barbieri:1995uv}. The framework involves a U(2)$_Q \times$ U(2)$_u \times$ U(2)$_d$ flavor symmetry, broken by spurions $\Delta Y_u$ in the $({\bf 2}, {\bf {\overline 2}}, {\bf 1})$ representation and $\Delta Y_d$ in the $({\bf 2}, {\bf 1}, {\bf {\overline 2}})$. Further spurions can allow mixing with the third generation. The choice that works best explaining the quark masses and mixings is a doublet $V$ in the $({\bf 2}, {\bf 1}, {\bf 1})$ representation. Because of the symmetry structure of the spurions, the dominant off-diagonal terms are entirely in $\delta_{LL}$, with terms in $\delta_{RR}$ further suppressed by small Yukawas. A recent update on the constraints on this scenario, including renormalization group effects that modify the flavor structure at low energies when U(2)$^3$ is imposed at a high scale, appeared in Ref.~\cite{Blankenburg:2012ah}. It found viable regions of parameter space without tachyons, but worked strictly within the MSSM, so needed relatively large $A$-terms and was not completely natural.

To sum up, if we wish to split the first and second generations from the third, the best route appears to be to impose the U(2)$^3$ flavor symmetry as studied in Ref.~\cite{Barbieri:2011ci} and its successors. Although we work with high-scale SUSY breaking, we see no reason that such a symmetry could not hold to good approximation. Approximate global symmetries (or exact discrete gauge symmetries) are known to arise in certain string constructions and need not be broken severely by generic Planck-suppressed operators. To the extent that there is some tension in the flavor sector related to splitting the first and second generation squarks from the third, it is shared in our model and in other incarnations of natural SUSY. 

On the other hand, the flavor problems associated with split generations could be avoided if all squarks are light but have somehow been hidden from detection so far, e.g. via $R$-parity violating decays~\cite{Barbier:2004ez,Csaki:2011ge,Graham:2012th} or decays to a hidden sector~\cite{Strassler:2006qa,ArkaniHamed:2008qp,Baumgart:2009tn,Fan:2011yu}. In these cases, the SUSY-breaking mechanism could be simple and flavor-blind, although model-building problems may remain.

It is clear that the model proposed in this paper is not the whole story. A complete picture of natural SUSY necessarily involves either a flavor model like U(2)$^3$ or a mechanism for hiding superpartners like $R$-parity violation. Both require detailed confrontation with direct and indirect constraints. Such studies are a vital part of understanding the status of natural SUSY, but the fact that such very different approaches could work shows that they are orthogonal to the problem of understanding the Higgs sector. A core requirement of natural SUSY is lifting the Higgs mass to 125 GeV, and in this paper we have discussed a way to do this with fewer new mass scales put in by hand than in most alternative models.

\section{Discussion}
\label{sec:discuss}

Finding a model to accommodate current constraints on supersymmetry while retaining naturalness is surprisingly challenging. The major challenge is to allow for a new quartic term, most readily accommodated by a singlet or a new $D$ term. Either of these possibilities generally entails a new low scale, with a questionable coincidence with the supersymmetry breaking scale.

In this paper, we've considered what is perhaps one of the more minimal ways to address this issue. We take the low scale seriously and assume it is associated with a composite sector. We furthermore relate the scale of compositeness to the fundamental supersymmetry-breaking scale. This allows us to address the Higgs sector.

On top of the Higgs bounds and possible hints, constraints on supersymmetric partners are also becoming quite stringent. An attractive way around the bounds is to have only the third generations squarks light, the stop in particular. This is readily accommodated in a geometric setting, or any model in which the top interacts less directly with the supersymmetry-breaking sector.

Even this is not completely flexible, however, as renormalization group constraints imply that the gluino will be at most about a factor of two heavier. The first and second generation squarks and right-handed sbottom must either be quite heavy (above about 6 TeV) even in models with favorable flavor symmetry structure, or must be hidden (e.g. through $R$-parity violating decays). Fully working out the constraints on the first and second generation squarks and how they have evaded direct and indirect bounds would be an interesting extension of this work.

Although constraining for models, this does mean that natural supersymmetric scenarios, particularly of this sort, will be tested at the LHC.  Meanwhile it is best to consider all natural possibilities to ensure that if such a scenario exists, we do find it. And if we don't, we will know the fate of weak scale supersymmetry.

\section*{Acknowledgments}
MR is supported by the Fundamental Laws Initiative of the Harvard Center for the Fundamental Laws of Nature. L.R. was supported in part by the the NSF grant PHY-0855591. We thank the Aspen Center for Physics for hospitality while a portion of this work was completed. We also thank an anonymous referee for pointing us to interesting recent flavor bounds on models with split generations.

\appendix

\section{SO(n) model: basics}
\label{app:SOnbasics}

As we have mentioned, in an SO($n$) theory with $n_f$ flavors, the conformal window is $\frac{3}{2}\left(n - 2\right) \leq n_f \leq 3\left(n-2\right)$~\cite{Intriligator:1995id}. Let's work out the one-loop estimate for the ratio between the $X$ mass and the confinement scale, as a function of $n$ and the number of flavors we start with. The NSVZ beta function is $\propto 3 \mu({\rm Adj}) - \sum_i \mu(i) (1 - \gamma_i)$, where $\mu({\bf r})$ is the Dynkin index of the representation ${\bf r}$ and $\gamma$ is the anomalous dimension. For SO($n$), at least in one choice of normalization, the Dynkin index of the fundamental is $\mu(\Box) = 2$ and of the adjoint is $\mu({\rm Adj}) = 2n-4$, leading to an anomalous dimension
\beq
\gamma = \frac{n_f - 3 n + 6}{n_f}.
\eeq
Crudely, we can estimate the gauge coupling at the fixed point by comparing this to the one-loop anomalous dimension:
\beq
\gamma = -\frac{2 g^2 C_2(\Box)}{16\pi^2}
\eeq
We can use $C_2(\Box) = \mu(\Box) \frac{{\rm dim}({\rm Adj})}{{\rm dim}(\Box)} = n-1$ to conclude that the fixed point value, in the 1-loop approximation to $\gamma$, is:
\beq
g_*^2 = \frac{8\pi^2}{n-1} \left(\frac{3(n-2)}{n_f} - 1\right).
\label{eq:fixedpointestimate}
\eeq
In this form, we find $g_* = 0$ at the upper end of the window, $n_f = 3(n-2)$, as expected. At the lower end of the window, we estimate $g_*^2 \left(n-1\right) = 8\pi^2$, which is clearly not a reliable calculation. Still, it's a rough guideline to where we have a strongly coupled fixed point; e.g., fixing the 't Hooft coupling $g^2_*(n-1) = \pi^2$ corresponds to a choice for the number of flavors of $n_f = \frac{8}{3}\left(n-2\right)$. So, roughly, anything an order-one fraction of the conformal window below $n_f = 3(n-2)$ will be at strong coupling.

Assuming that we started at a fixed point with $n_f$ flavors, we can ask what the confinement scale will be if we abruptly integrate out some flavors at a scale $M$ and reduce to a theory with $n-4$ flavors. In the approximation that $\gamma$ remains fixed at its fixed point value,\footnote{It won't, of course, but this could capture slightly more of the physics than the most naive one-loop estimate of the beta function that treats the quarks as free fields.} we have a beta function which is now given at one loop by
\beq
\beta(g^2) \approx \frac{-6g^4 }{8\pi^2} (n-2) \left(1 - \frac{n-4}{n_f}\right),
\eeq
so defining $b_0 \equiv 6(n-2)\left(1 - \frac{n-4}{n_f}\right)$, we have a confinement scale
\beq
\Lambda \approx M \exp\left(-\frac{8\pi^2}{b_0 g_*^2}\right) \approx M \exp\left(\frac{-(n-1)n_f^2} {6(n-2)(3n - n_f - 6)(n_f + 4 - n)}\right),
\eeq
where we have used the estimate~\ref{eq:fixedpointestimate} for $g_*$ in the second step. For example, if we fix $n = 10$ and $n_f = 20$, which is still near enough to the upper end of the conformal window that the estimate for $g_*$ is not completely unreliable, this crude estimate gives $\Lambda \approx M/4$. All of this serves as a basic sanity check: we can see in an approximately controlled way that an SO($n$) gauge theory can generate a confinement scale of order, but somewhat below, $m_{3/2}$ after we use the Giudice-Masiero mechanism to integrate out some flavors.

\section{Loops for the $S$ linear terms}
\label{app:loops}

To understand the linear term in the potential, as well as the tadpole, let's compute an effective K\"ahler potential. We start by writing the superpotential including effective $\chi{\tilde \chi}$ and $\sigma {\tilde \sigma}$ mass terms:
\beq
\int d^2 \theta~\left[M_\chi \chi {\tilde \chi} + M _\sigma \sigma {\tilde \sigma} + y \left(\phi \chi \sigma + \phi {\tilde \chi} {\tilde \sigma}\right)\right],
\eeq
where the Giudice-Masiero masses are parametrized by
\beq
M_{\chi,\sigma} = \mu_{\chi,\sigma} + \theta^2 B_{\chi,\sigma}.
\eeq
(Here we assume that the fields have been canonically normalized, so wavefunction factors from strong dynamics are incorporated into the $M$ and $y$ values. Wavefunction overlaps from integrating out an extra dimension are also absorbed into these factors.) Then the effective K\"ahler potential for $\phi$ is given in terms of the mass matrix ${\cal M}$ for the chiral superfields $\sigma$ and $\chi$ by~\cite{Grisaru:1996ve}:
\beq
K_{\rm eff} = -\frac{1}{32\pi^2} {\rm Tr}\left( {\cal M}{\cal M}^\dagger \left( \log \frac{{\cal M}{\cal M}^\dagger}{\Lambda^2} - 1 \right) \right).
\eeq
(For the effective K\"ahler potential to be reliable, we should assume $B_{\chi,\phi} \ll \mu_{\chi,\phi}^2$.) Turning the crank we find that this generates terms
\beq
\int d^4\theta \frac{y^2}{32\pi^2} \frac{M_\chi^\dagger M_\phi^\dagger}{\left|M_\chi\right|^2 - \left|M_\phi\right|^2} \log\left|\frac{M_\phi}{M_\chi}\right|^2 \phi^2 + c.c.
\label{eq:Keffphi2}
\eeq
which we can interpret as containing both a linear tadpole term in the potential, $\int d^4 x T S$, as well as a term in the superpotential $\int d^2\theta f S$. In particular, writing $\mu_\phi = \mu, \mu_\chi = \xi \mu$, $B_\phi = B, B_\chi = \eta B$, and using $\phi^2 \approx \frac{\sqrt{n}}{4\pi} \Lambda S$, we have:
\beqs
f & \approx & \frac{\sqrt{n}}{4\pi}\frac{y^2}{32\pi^2} \frac{B\Lambda}{\mu} g_f(\xi,\eta),\label{eq:fterm}\\
T & \approx & \frac{\sqrt{n}}{4\pi}\frac{y^2}{32\pi^2} \left(\frac{B^2\Lambda}{\mu^2} g_T(\xi,\eta) +  \frac{BF_\Lambda}{\mu} g_f(\xi,\eta)\right),\label{eq:tadpole}
\eeqs
where $g_f(\xi,\eta) = \frac{-2 (\xi^3 - \eta)\log\xi - (\xi^2-1)(\eta-\xi)}{(\xi^2-1)^2}$ and $g_T(\xi,\eta) = \frac{(\xi^4-1)(\eta-\xi)^2-2\xi(2\eta^2 \xi+2\xi^3+\eta(\xi^2+1)^2)\log\xi}{\xi(\xi^2-1)^3}$ are simple numerical functions of the ratios among the various Giudice-Masiero scales, which we expect to be order 1. (In the limit $\xi,\eta \to 1$, they reduce to $g_f \to -1$ and $g_T \to 1$.) The part of $T$ proportional to $F_\Lambda$ arises from reading off a ${\bar \theta}^2$ component from the coefficient of $\phi^2$ in Eq.~\ref{eq:Keffphi2} and a $\theta^2$ component from the factor of $\Lambda$ that arises in converting $\phi^2$ to $S$. Because $g_T$ and $g_f$ have opposite sign, in the limit that $B/\mu$ is identical for all of the fields $\Phi$, $\chi$, and $\sigma$, $T$ will vanish. This is similar to the vanishing $A$-term remarked upon surrounding Eq.~\ref{eq:Flambda}. Thus, $T$ is naturally proportional to a {\em difference} of $B/\mu$ terms, $\delta B/\mu$. However, because $\sigma$ is not charged under SO($n$), it is reasonable for it to have a different $B/\mu$ ratio from that of $\chi$ and $\Phi$, so we expect $T/f$ to be typically larger than $A_\lambda/\lambda$.

We {\em also} generate a term $\propto \phi^\dagger \phi$ which may be interpreted as a soft mass for the scalars making up $S$:
\beq
\int d^4\theta \frac{y^2}{16\pi^2}\frac{\left|M_\phi\right|^2 \log\left|M_\phi\right|^2 -\left|M_\chi\right|^2 \log\left|M_\chi\right|^2}{\left|M_\chi\right|^2 - \left|M_\phi\right|^2} \phi^\dagger \phi.
\eeq
(The ambiguity in the scale of the logarithm simply corresponds to a supersymmetric wavefunction renormalization of $\phi$.) The resulting $\phi$ soft mass,
\beq
m_\phi^2 = \frac{y^2}{16\pi^2} \frac{B^2}{\mu^2} \frac{(\eta-\xi)^2 \left(2(\xi^2-1)-(\xi^2+1)\log\xi^2\right)}{(\xi^2-1)^3},
\eeq
is small enough that it really should be interpreted as perturbing the confining theory. In the limit $\xi, \eta \to 1$, it vanishes. For general $\xi,\eta$, it could be a soft term of order a few hundred GeV. Unlike a $\phi\phi$ term, we can't directly express it in terms of $S$, but since it splits the scalar and fermion in $\phi$, it will in turn split the scalar and fermion bound states in $S$. In other words, we can model this by assuming an electroweak-scale soft mass for $S$ itself. We expect its effect to be subdominant relative to the tadpole.

\section{Removing unwanted mesons}
\label{app:removal}

We expect the scalar fields in these mesons to obtain SUSY-breaking masses, through anomaly mediation if nothing else. However, the fermions (mesinos) can be light and thus problematic for BBN. One way to address this problem would be to weakly gauge an SU(3) subgroup of the flavor symmetry acting on the SO($n$) fundamentals; we take $\phi$ to be an SU(3) singlet and group the remaining six light flavors into a fundamental and antifundamental of SU(3). If the SU(3) group confines at a scale below the SO($n$) confinement scale, but above the BBN temperature, most of the unwanted mesons will gain mass and decouple before BBN (because they will fall into a ${\bf 3},~{\bf \overline{3}},~{\bf 6},~{\bf \overline{6}}$ and ${\bf 8}$ of SU(3)). Two singlets remain, one of which is our $S$ field, and one of which is an extra mesino. One extra Weyl fermion at BBN is still a possibility allowed by data. (Essentially, the extra mesino is like a sterile neutrino.)

Another approach is to add higher-dimension operators that can give various fields a small mass. If the superpotential contains $\frac{1}{\Lambda} q_i q_j q_k q_l$, for instance, in the low-energy theory after SO($n$) confinement this becomes an effective meson mass. Such terms would need to break enough flavor symmetries to give masses to all the mesons, and the scale $\Lambda$ would need to be at or below about $10^{10}~{\rm GeV}$ to make the mesons heavy enough to not be problematic for BBN. Perhaps this scale could be related to other interesting physics like Peccei-Quinn breaking or the scale $\sqrt{F}$.

Because our main goal was to illustrate some of the physics resulting from the choice of making all low-energy scales relate to $m_{3/2}$, we will not dwell on these model-building details.

\end{document}